\documentclass[letter,twocolumn,showpacs,aps,prb,floatfix]{revtex4}
\usepackage{graphicx,subfigure,epsfig,verbatim,psfrag,amsmath,amssymb,color}
\newcommand{\red}[1]{{\textcolor{red}{#1}}}
\newcommand{\blue}[1]{{\textcolor{blue}{#1}}}
\newcommand{\mean}[1]{\left<#1\right>}

\makeatletter
\input epsf

\def\be{\begin{equation}}
\def\ee{\end{equation}}
\def\ba{\begin{eqnarray}}
\def\ea{\end{eqnarray}}
\makeatother

\newcommand{\up}{\uparrow}
\newcommand{\dn}{\downarrow}

\newcommand{\Tr}{\mathop{\text{Tr}}}

\begin{document}

\title{XXZ and Ising Spins on the Triangular Kagome Lattice} 
\author{Dao-Xin Yao, Y.~L.~Loh, and E.~W.~Carlson}
\affiliation{Department of Physics, Purdue University, West Lafayette, IN  47907}
\author{Michael Ma}
\affiliation{Department of Physics, University of Cincinnati, Cincinnati,
OH  45221}
\date{\today}

\begin{abstract}
The recently fabricated two-dimensional magnetic materials
$\mbox{Cu}_{9}\mbox{X}_2(\mbox{cpa})_{6}\cdot x\mbox{H}_2\mbox{O}$
(cpa=2-carboxypentonic acid; X=F,Cl,Br) have copper sites which form a
triangular kagome lattice (TKL), formed by introducing small triangles
(``$a$-trimers'') inside of each kagome triangle (``$b$-trimer'').  We
show that in the limit where spins residing on $b$-trimers have Ising
character, quantum fluctuations of XXZ spins residing on the
$a$-trimers can be exactly accounted for in the absence of applied
field.  This is accomplished through a mapping to the kagome Ising
model, for which exact analytic solutions exist.  We derive the
complete finite temperature phase diagram for this XXZ-Ising model,
including the residual zero temperature entropies of the seven ground
state phases. Whereas the disordered (spin liquid) ground state of the
pure Ising TKL model has macroscopic residual entropy
$\rm{ln}72=4.2767...$ per unit cell, the introduction of transverse
(quantum) couplings between neighboring $a$-spins reduces this entropy
to $2.5258...$ per unit cell.  In the presence of applied magnetic
field, we map the TKL XXZ-Ising model to the kagome Ising model with
three-spin interactions, and derive the ground state phase diagram. A
small (or even infinitesimal) field leads to a new phase that
corresponds to a non-intersecting loop gas on the kagome lattice, with
entropy $1.4053...$ per unit cell and a mean magnetization for the $b$-spins of $0.12(1)$ per site. In addition, we find that for moderate
applied field, there is a critical spin liquid phase which maps to
close-packed dimers on the honeycomb lattice, which survives even when
the $a$-spins are in the Heisenberg limit.
\end{abstract}
 \maketitle


\section{Introduction}

Geometrically frustrated spin systems hold promise for finding new
phases of matter, such as classical and quantum spin liquid ground
states.  Of considerable interest has been the discovery of a stable
phase with deconfined spinons in a model of quantum dimers on the
(geometrically frustrated) triangular lattice, {\em i.e.} a spin
liquid.\cite{moessner2001} Beyond the interest in fundamental
theoretical issues of these
models,\cite{moessner2000,moessner2001,moessner2003} physical
realizations of these systems may have technological applications in
achieving lower temperatures through adiabatic demagnetization.  Since
such techniques require a material which can remain in a disordered,
paramagnetic state to very low temperatures (rather than undergoing a
phase transition to an ordered state), this makes geometrically
frustrated spin systems attractive for such applications.

\begin{figure}[t]
\includegraphics[width=0.6\columnwidth]{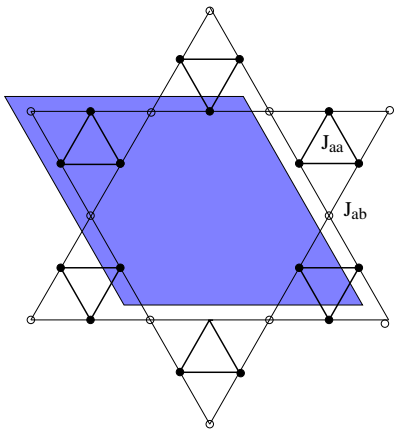}
\caption{(Color online) The triangular kagome lattice (TKL) with the
nearest-neighbor couplings $J_{aa}$ and $J_{ab}^Z$. Solid circles are
``a'' sites and open circles are ``b'' sites. The shaded region
represents the unit cell.} 
\label{f:tkl_lattice}
\end{figure}

Recently, a new class of two-dimensional magnetic materials
$\mbox{Cu}_{9}\mbox{X}_2(\mbox{cpa})_{6}\cdot x\mbox{H}_2\mbox{O}$
(cpa=2-carboxypentonic acid, a derivative of ascorbic acid; X=F,Cl,Br)
~\cite{gonzalez93,maruti94, mekata98} was fabricated in which Cu spins
reside on a triangular kagome lattice (TKL), formed by inserting an
extra set of triangles ($a$-trimers) inside of the kagome triangles
($b$-trimers).  (See Fig.~\ref{f:tkl_lattice}.)  In a recent
paper\cite{lohyaocarlson2007}, we analyzed the thermodynamic behavior
of Ising spins on this lattice using exact analytic methods as well as
Monte Carlo simulations in finite field.  In this paper, we extend our
analysis to include quantum fluctuations of the spins on $a$-trimers,
{\em i.e.} we study an XXZ-Ising model on the TKL.  The Cu spins in
$\mbox{Cu}_{9}\mbox{X}_2(\mbox{cpa})_{6}\cdot x\mbox{H}_2\mbox{O}$
likely have isotropic Heisenberg interactions, for which exact
solutions are currently inaccessible on a frustrated lattice.
However, in the limit where spins on $b$-trimers have Ising character,
it is still possible to take into account the quantum fluctuations on
$a$-trimers {\em exactly} through a mapping to the kagome Ising model.
In the presence of applied field, the model maps to the kagome Ising
model with three-spin interactions.  We present exact results for the
phase diagram at all temperatures without applied field, and at zero
temperature in the presence of applied field.

In the absence of applied field, the zero temperature phase diagram is
richer than the case where both $a$ and $b$ spins are in the Ising
limit.  The ordered phase survives quantum fluctuations, but for large
enough transverse coupling, there is a first order transition to a
state with lower total spin on the $a$-trimers.  For the disordered
phase, the quantum fluctuations of the $a$-spins partially lift the
ground state degeneracy. In the presence of applied field, the phase
diagram is even more rich. The ``honeycomb dimer'' phase, which is
present in applied field for the purely Ising version of the
model,\cite{lohyaocarlson2007} survives the introduction of quantum
fluctuations.  However, rather than arising for infinitesimally
applied field, a finite field is now required. Infinitesimal applied
field in the presence of quantum fluctuations of the $a$ spins results
in a new phase, which we have mapped to a non-intersecting loop gas on
the kagome lattice.

This paper is organized as follows.  In Section~\ref{model} we
introduce the XXZ-Ising model on the TKL. In Section~\ref{s:zero-field} we
present an exact mapping to the kagome Ising model, and 
we derive the finite temperature phase diagram in zero field.
In Section~\ref{finiteh}, we study the finite field case,
using an exact mapping to
the kagome Ising model in a field with a three-spin
coupling term.
In Section~\ref{conclusions} we present our 
discussion and conclusions.

\section{XXZ-Ising model on the TKL\label{model}}

The TKL can be obtained by inserting 
an extra set of 
triangles inside of the triangles of
the kagome lattice.  Alternatively, it can be derived from the
triangular lattice by periodically deleting seven out of every sixteen
lattice sites.  This structure has two different spin sublattices,
$a$ and $b$, which correspond to small trimers and large trimers,
respectively.  Each spin has four nearest neighbors.  The unit cell
contains a total of 9 spins  (6 on the $a$ sublattice, and 
3 on the $b$ sublattice).
The  space group of the TKL is the same as that of the
hexagonal lattice, $p6m$, in Hermann-Mauguin notation.  
The shaded region in 
Fig.~\ref{f:tkl_lattice} encompasses one unit cell of the TKL.

The Cu spins in  real materials have $S=1/2$,
and quantum effects cannot be neglected {\em a priori}.
The Ising limit of this model has been previously
considered by Zheng and Sun\cite{zheng05},
as well as by three of us\cite{lohyaocarlson2007}.
The ground state phase diagram\cite{zheng05,lohyaocarlson2007}
as well as many experimentally testable thermodynamic quantities\cite{lohyaocarlson2007}
have been calculated.
In this paper, we include the quantum fluctuations of the
$a$ spins ({\em i.e.} those on $a$-trimers),
while treating the $b$ spins as classical Ising spins.
In this limit, the $a$ spins can be integrated out exactly,
leaving an Ising model of the $b$ spins on a 
kagome lattice, which we solve exactly.

We consider a
model in which the exchange coupling between neighboring $a$-spins is of
the XXZ type 
and the coupling between
neighboring $a$- and $b$-spins is Ising-like. 
Whereas this treats the $a$-spins as fully quantum mechanical,
the $b$-spins are classical.
The Hamiltonian is 
        \begin{align}
        H
        &= 
        -\sum_{\mean{i\in a,j\in a}}  
        \left[
        J_{aa}^Z S_i^Z S_j^Z + J_{aa}^X (S_i^X S_j^X + S_i^Y S_j^Y)
        \right]
\nonumber\\&{}
        -\sum_{\mean{i\in a,j\in b}}  
        J_{ab}^Z S_i^Z S_j^Z   - h \sum_{i} S_i^z 
        \label{e:tkl-quantum-heisenberg-hamiltonian}
        \end{align}
where $S_i^{X,Y,Z}$ are the $S=1/2$ spin operators at site $i$, angle
brackets indicate summations over nearest neighbors, and $h$ is an
external magnetic field.  
With this sign convention, positive coupling $J>0$ corresponds
to ferromagnetic interactions, and negative coupling $J<0$ 
is antiferromagnetic.
This model contains
the following energy scales as parameters: $J_{aa}^Z$, $J_{aa}^X$,
$J_{ab}^Z$, $T$, and $h$.  We will take $|J_{ab}^Z|$ as the unit of
energy.  


\begin{figure}[tb]
\psfrag{Jaa}{\red{$J_{aa}^{Z,X}$}}
\psfrag{Jab}{\blue{$J_{ab}^Z$}}
\psfrag{Sa1}{\red{$S_{a1}$}}
\psfrag{Sa2}{\red{$S_{a2}$}}
\psfrag{Sa3}{\red{$S_{a3}$}}
\psfrag{Sb1}{\blue{$S_{b1}$}}
\psfrag{Sb2}{\blue{$S_{b2}$}}
\psfrag{Sb3}{\blue{$S_{b3}$}}
\includegraphics[width=0.6\columnwidth]{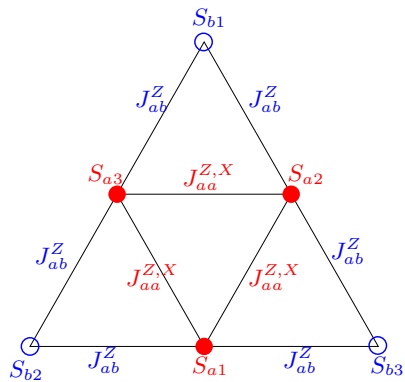}
\caption{(Color online) A hexamer consisting of three $a$-spins (filled circles) and three $b$-spins (open circles).  The coupling between
$a$-spins is of the XXZ type, whereas the $a$--$b$ coupling is of
Ising type.  The mixed Heisenberg-Ising model on the TKL can be
fruitfully analysed by breaking it up into hexamer units.}
\label{f:hexamer}
\end{figure}

When  $J_{aa}^X = 0$, this model reduces to the Ising model on
the TKL, which was studied exhaustively in
Ref.~\onlinecite{lohyaocarlson2007}.  (See also Ref.~\onlinecite{zheng05}.)
The $J_{aa}^X$ terms permit transverse quantum 
fluctuations within each $a$-trimer.  However, the quantum fluctuations
are confined to reside within the $a$-trimers, 
as can be seen by the fact that the total $S^z$ of each $a$-trimer is a conserved quantity.

In spite of the frustrated nature of the model,  it turns out that it can be 
exactly
solved at all temperatures for $h=0$ and 
at zero temperature for finite $h$.
The Hamiltonian can be  written as a sum over hexamers,
(see Fig.~\ref{f:hexamer})
where each hexamer consists of an $a$-trimer and its enclosing $b$-trimer,
$\hat H = \sum_n \hat H_n$. 
The Hamiltonian for each hexamer is given by   
        \begin{align}
        \hat H_n &= - \sum_{\mean{ij}} \left[ J_{aa}^Z S_{ai}^Z
        S_{aj}^Z + J_{aa}^X (S_{ai}^X S_{aj}^X + S_{ai}^Y S_{aj}^Y)
        \right] \nonumber\\&{} - \sum_{i} \sum_{j\neq i} 
        J_{ab}^Z S_{ai}^Z S_{bj}^Z  - h \sum_{i}S_{ai}^Z-\frac{h}{2}\sum_{j}S_{bj}^Z~.
        \label{e:hexamer-hamiltonian}
        \end{align}
Note the factor of  ``$\frac{h}{2}$''
in front of $S_{bj}^Z$, which is required because each $b$-spin is shared by two hexamers. 

Different $\hat H_n$ commute with each other.  Furthermore, each $a$-spin appears in one $\hat H_n$ but no others. Therefore, we can perform the trace over all of the $a$-spins to give an effective Hamiltonian involving only $b$-spins as follows: 
        \begin{align}
        Z &=\Tr \exp (-\beta \hat H) \\
        & =\sum_{\{S_b\}} \Tr_{S_{a1},S_{a2},S_{a3}} \exp \left[-\beta \hat H (S_b,S_a )\right] \\
        & =\sum_{\{S_b\}} e^{-\beta \hat H_\text{eff}(S_b)}.
        \label{e:partition-function-factorization}
        \end{align}

The contribution to the partition function coming from the  trace over the $a$-spins within each hexamer 
depends on the values of the surrounding $b$-spins, i.e.,
        \begin{align}
        Z(S_{b1},S_{b2},S_{b3})
        &= \Tr_{S_{a1},S_{a2},S_{a3}} \exp \left[-\beta \hat H_n (S_b, \hat S_a)\right]
        \end{align}

The trace can be evaluated by diagonalizing the hexamer Hamiltonians
for all eight configurations of the enclosing $b$-spins,
in order to obtain 
the energy eigenvalues
$E_j(S_{b1},S_{b2},S_{b3},h)$, $j=1,2,3,\dotsc,8$ (see   
Fig.~\ref{f:energy-levels-of-a-trimer}), 
and subsequently calculating
$ Z(\{S_b\}) = \sum_j \exp\left[-\beta E_j(\{S_b\})\right]$. 
The energy diagonalization is particularly simple since the total $S^Z$ of each $a$-trimer is a good quantum number. In the case where the surrounding $b$-spins are in the $(\up\up\up)$ or $(\dn\dn\dn)$ configurations, total $S^2$ is also conserved.
Due to the local $C_3$ symmetry of each hexamer,
${E_j(\up\up\dn,h)}={E_j(\up\dn\up,h)}={E_j(\dn\up\up,h)}$.
In addition, the energy eigenvalues respect time-reversal symmetry, 
so that ${E_j(\dn\dn\dn,h)}={E_j(\up\up\up,-h)}$,
and ${E_j(\up\dn\dn,h)}={E_j(\dn\up\up,-h)}$.

\begin{widetext} 
\begin{center} 
\begin{table}[htbp]
\begin{tabular}{|c|c|} 
\hline
  $4E_j(\up\up\up,h)$ & $4E_j(\up\up\dn,h)$      \\ \hline
$-4 J_{aa}^X+J_{aa}^Z-2
  J_{ab}^Z-5h$ & $2  J_{aa}^X+J_{aa}^Z-2 J_{ab}^Z- 3h $ \\
$2 J_{aa}^X+J_{aa}^Z-2 J_{ab}^Z-5h$ & $-3 J_{aa}^Z-2 J_{ab}^Z-7h$ \\
$2 J_{aa}^X+J_{aa}^Z-2 J_{ab}^Z-5h$ & $2 J_{ab}^Z-3 J_{aa}^Z+5h$ \\
$-3J_{aa}^Z-6 J_{ab}^Z-9h$ & $2 J_{aa}^X+J_{aa}^Z+2 J_{ab}^Z+h$ \\
$-3J_{aa}^Z+6 J_{ab}^Z+3h$ & $-J_{aa}^X+J_{aa}^Z-\sqrt{-4 J_{ab}^Z J_{aa}^{X  }+9 J_{aa}^{X 2 }+4 J_{ab}^Z {}^2}+h$ \\
$-4 J_{aa}^X+J_{aa}^Z+2 J_{ab}^Z-h$ & $-J_{aa}^X+J_{aa}^Z+\sqrt{-4
   J_{ab}^Z J_{aa}^X+9 J_{aa}^{X 2 }+4 J_{ab}^Z {}^2}+h$ \\
$2 J_{aa}^X+J_{aa}^Z+2 J_{ab}^Z-h$ & $-J_{aa}^X+J_{aa}^Z-\sqrt{4 J_{ab}^Z
   J_{aa}^X+9 J_{aa}^{X 2 }+4 J_{ab}^Z {}^2}-3h$ \\
 $2 J_{aa}^X+J_{aa}^Z+2 J_{ab}^Z-h$ & $-J_{aa}^X+J_{aa}^Z+\sqrt{4 J_{ab}^Z
   J_{aa}^X+9 J_{aa}^{X 2 }+4 J_{ab}^Z {}^2}-3h$
\\ \hline
\end{tabular}
\caption{The eight energy levels of a hexamer arising from the $a$-spin degrees of freedom, for two configurations of the enclosing $b$-spins ($\up\up\up$ and $\up\up\dn$).
\label{t:hexamer-energy-levels}
}
\end{table}
\end{center} 

Hence we find
\begin{align}
Z(\up\up\up,h) &= 
2 e^{\frac{1}{4} (h-2 J_{aa}^X-J_{aa}^Z-2 J_{ab}^Z) \beta
}+e^{\frac{1}{4} (h+4 J_{aa}^X-J_{aa}^Z-2 J_{ab}^Z)   \beta }+2
e^{\frac{1}{4} (5 h-2 J_{aa}^X-J_{aa}^Z+2 J_{ab}^Z) \beta } \nonumber \\& {}
+e^{\frac{1}{4} (5 h+4J_{aa}^X-J_{aa}^Z+2 J_{ab}^Z) \beta } 
+e^{-\frac{3 h \beta }{4}+\frac{3 J_{aa}^Z \beta }{4}-\frac{3 J_{ab}^Z
   \beta }{2}}+e^{\frac{9 h \beta }{4}+\frac{3 J_{aa}^Z \beta }{4}+\frac{3
   J_{ab}^Z \beta }{2}} \label{e:zuuu-and-zduu-1}\\
Z(\up\up\dn,h) &= 
e^{-\frac{1}{4} (5 h-3 J_{aa}^Z+2 J_{ab}^Z) \beta }+e^{\frac{1}{4} (3 h-2 J_{aa}^X-J_{aa}^Z+2 J_{ab}^Z) \beta
   }+e^{-\frac{1}{4} (h+2 J_{aa}^X+J_{aa}^Z+2 J_{ab}^Z) \beta }+e^{\frac{1}{4} (7 h+3 J_{aa}^Z
+2 J_{ab}^Z) \beta  }  \nonumber \\& {}
+e^{\frac{1}{4} \left(-h+J_{aa}^X-J_{aa}^Z+\sqrt{9 J_{aa}^X {}^2-4 J_{ab}^Z J_{aa}^X+4 J_{ab}^Z {}^2}\right)
   \beta }+e^{-\frac{1}{4} \left(h-J_{aa}^X+J_{aa}^Z+\sqrt{9 J_{aa}^X {}^2-4 J_{ab}^Z J_{aa}^X+4
   J_{ab}^Z {}^2}\right) \beta } \nonumber \\& {}
+e^{\frac{1}{4} \left(3 h+J_{aa}^X-J_{aa}^Z+\sqrt{9 J_{aa}^X {}^2+4 J_{ab}^Z
   J_{aa}^X+4 J_{ab}^Z {}^2}\right) \beta }+e^{-\frac{1}{4} \left(-3 h-J_{aa}^X+J_{aa}^Z+\sqrt{9 J_{aa}^X {}^2+4
   J_{ab}^Z J_{aa}^X+4 J_{ab}^Z {}^2}\right) \beta }     \\
Z(\dn\dn\dn,h) &= Z(\up\up\up,-h)  \\
Z(\up\dn\dn,h) &= Z(\up\up\dn,-h)~.
\label{e:zuuu-and-zduu}
\end{align}
\end{widetext}


\section{Zero Field \label{s:zero-field}}
\subsection{Exact mapping to the kagome Ising model \label{sec:exact}}  
When $h=0$, the trace over $a$-spins maps the XXZ-Ising model to the ferromagnetic Ising model on
the kagome lattice exactly, so that up to a temperature-dependent additive constant, we have
 \begin{equation}
        H_\text{eff}(\{\sigma_b\})=- J_{bb}\sum_{\mean{i,j}} \sigma_{bi} \sigma_{bj},
\end{equation}
where $J_{bb}$ is the effective $b$-spin coupling and $\sigma_{bi}=\pm 1$ for consistency with the Ising model
literature.

The simple form of $H_\text{eff}$ (two-spin nearest neighbor interactions only) can be understood easily. Since the hexamer Hamitonians commute between different hexamers, the trace over $a$-spins in a hexamer couples only the three $b$-spins in that hexamer. When $h=0$, $H_\text{eff}$ must have global up-down symmetry and so it  cannot contain odd powers of $\sigma$. Taking into account that $\sigma^2 =1$ then implies the quadratic nearest neighbor $H_\text{eff}$ above.  To find $J_{bb}$, we evaluate the  two different actions $Z(\up\up\up)$ and
$Z(\up\up\dn)$ (others are related by permutation and up down symmetry of the $b$-spins) and match them to
        \begin{align}
        Z(\sigma_{b1},\sigma_{b2},\sigma_{b3})
        &=Z_a~ \exp \beta J_{bb} (\sigma_{b2} \sigma_{b3} + \sigma_{b3} \sigma_{b1} + \sigma_{b1} \sigma_{b2})
        \end{align}
whereupon we obtain
        \begin{align}
        Z_a &= Z(\up\up\up) ^ {1/4} Z(\dn\up\up) ^{3/4}, \\
        \beta J_{bb} &= \frac{1}{4} \ln \frac{Z(\up\up\up)}{Z(\dn\up\up)}.
        \label{e:a-and-jbb}
        \end{align}

Thus, we see that integrating out the $a$-spins gives rise to effective
couplings between the $b$-spins, which corresponds to the classical Ising model
on the kagome lattice.  The effective Ising coupling $J_{bb}$ is a
complicated function of the original couplings $J_{aa}^{Z,X}$ and
$J_{ab}^Z$ , described by 
Eqs.~\eqref{e:zuuu-and-zduu-1}--\eqref{e:zuuu-and-zduu}
and Eq.~\eqref{e:a-and-jbb}, and of the temperature.  However, in all cases, it is non-negative and hence
the effective $b$-spin coupling is 
ferromagnetic.

The ferromagnetic kagome Ising model has an
exact solution, which has been known for some time\cite{syozi51}.
Therefore, the XXZ-Ising model on the TKL can also be solved
exactly.  In particular, the critical temperature, free energy, energy
density, entropy, and specific heat can be calculated in the same
manner as in Ref.~\onlinecite{lohyaocarlson2007}, which we outline
in the next section.

\begin{figure}[t]
{\centering
  \subfigure[~$J_{aa}^Z = -3$\label{f:Ej3}]{\includegraphics[width=0.95\columnwidth]{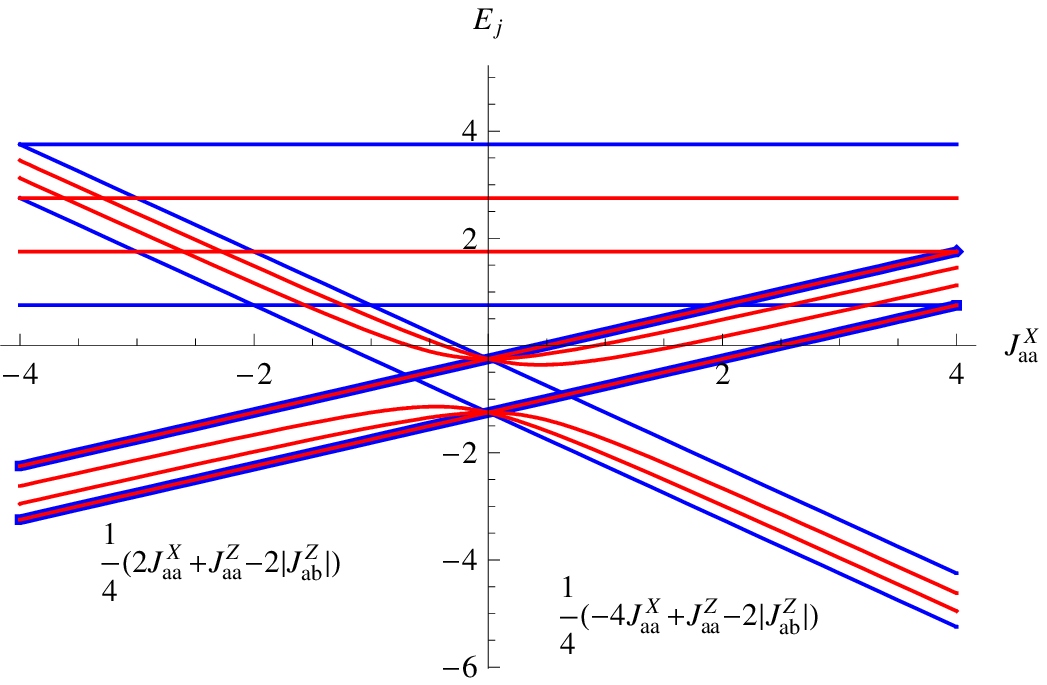}}
  \subfigure[~$J_{aa}^Z = +1$\label{f:Ej1}]{\includegraphics[width=0.95\columnwidth]{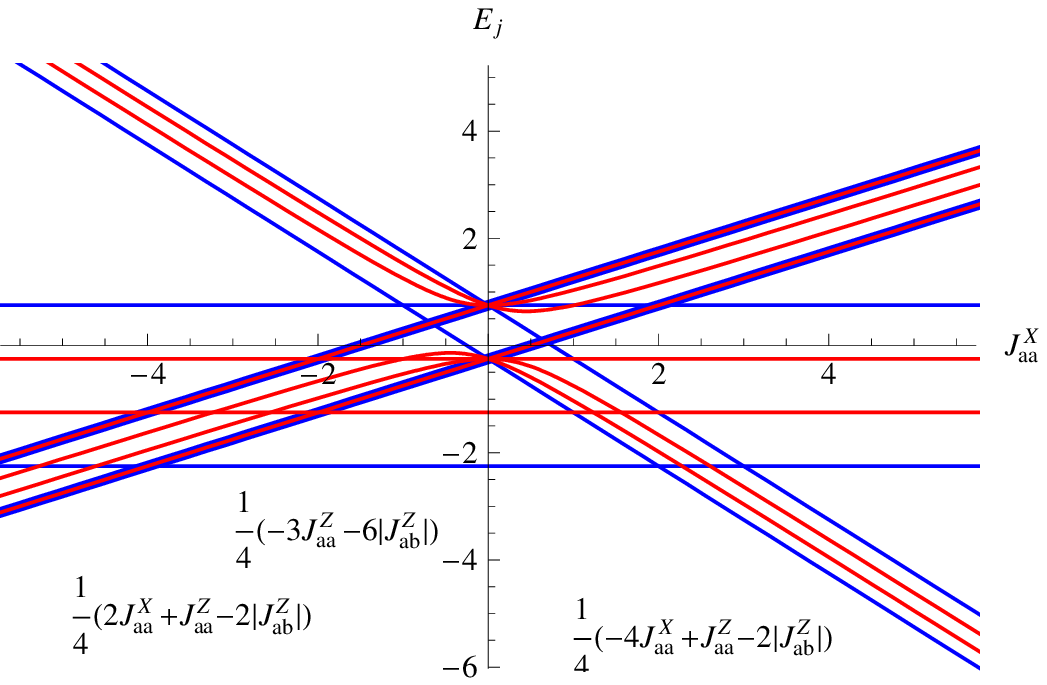}}
}
\caption{(Color online) Energy levels of the 
hexamers
when the enclosing $b$ spins are in the $(\up\up\up)$ 
configuration (blue lines)
and when they are in the $(\dn\up\up)$  configuration (red lines).
The energies are shown as a function 
of the transverse coupling $J_{aa}^X$, for $|J_{ab}^Z|= 1$ and for two different values of $J_{aa}^Z$.
(The energy levels are independent of the sign of $J_{ab}^Z$.)  
The thick blue-red-blue lines represent triply degenerate states,
which include two degenerate states from the $(\up\up\up)$
subspace of the $b$-spins, and one state from the $(\dn\up\up)$ subspace.
Panel (a) shows a representative phase transition from phase $I(1/2)$
to Phase IX for $J_{aa}^Z < -|J_{ab}^Z|$.
Panel (b) shows a representative set of phase transitions from phase $I(1/2)$
to $I(3/2)$ to phase IX for $J_{aa}^Z>-|J_{ab}^Z|$.
\label{f:energy-levels-of-a-trimer}}
\end{figure}

\subsection{Critical temperature, free energy, and entropy}

The kagome Ising model has a phase transition to a ferromagnetic ordered state
at a coupling strength $\beta J_\text{kagome}^{c} = \frac{1}{4} \ln (3
+ \sqrt{12}) = 0.466566...$.  Therefore the TKL XXZ-Ising
model 
has a  phase transition 
when $\beta J_{bb} (J_{aa}^Z, J_{aa}^X, J_{ab}^Z, \beta) = \beta
J_\text{kagome}^{c}$.  
The ferromagnetic state of the kagome model corresponds to 
a {\em ferro}magnetic phase in the TKL for $J_{ab}^Z>0$ (since then the $a$-spins
are aligned with the $b$-spins),
and to a  {\em ferri}magnetic phase for $J_{ab}^Z<0$ (since then the $a$-spins 
are antialigned with the $b$-spins).

For convenience, we define the free energy as $F=\ln Z$. Based on the discussion above, we see that the free energy of the TKL XXZ-Ising model is given by the
sum of the free energy of the kagome Ising ferromagnet and 
a term that arises from  integrating out the
$a$-trimers,  i.e.
        \begin{align}
        f(J_{aa}^Z, J_{aa}^X, J_{ab}^Z, \beta)
        &=f_b (\beta J_{bb}) + 2 f_a
        \label{e:total-free-energy}
        \end{align}
where $\beta J_{bb} (J_{aa}^Z, J_{aa}^X, J_{ab}^Z, \beta)$ is the
effective kagome coupling from Eq.~\ref{e:a-and-jbb}.
The free energy per  TKL unit cell is $f$; $f_b$ is the free energy per kagome unit cell; and $f_a = \ln
Z_a$ is the free energy
contribution per $a$-trimer. The TKL unit
cell corresponds to one kagome unit cell, and it contains two $a$-trimers,
hence the factors in the above equation.

Other thermodynamic quantities can be obtained by differentiating $f$ with respect to $\beta$.  It will be convenient to define the quantities
	\begin{align}
	U_j(\{\sigma_b\})
	&
	= \frac{1}{Z} \sum_j 
		E_j(\{\sigma_b\})
		e^{ -\beta E_j(\{\sigma_b\}) }
	,\\
	C_j(\{\sigma_b\})
	&
	=  \beta^2  \left(
		\frac{1}{Z} \sum_j 
		E_j(\{\sigma_b\})^2
		e^{ -\beta E_j(\{\sigma_b\}) }
		- U_j(\{\sigma_b\})^2
			\right)
	,
	\end{align}
where $\{\sigma_b\} = \up\up\up$ or $\up\up\dn$.
Then, the energy per unit cell is
	\begin{align}
	u
	&= -\frac{df}{d\beta} \\
	&=
		\frac{U_{\up\up\up} + 3 U_{\up\up\dn}}{2} 
	+	\frac{U_{\up\up\dn} - U_{\up\up\up}}{4}
		\widetilde{u}_\text{kag} (\beta J_{bb})
	,
	\end{align}
and the heat capacity per unit cell is
	\begin{align}
	c
	&= \frac{du}{dT} \\
	&=
		\frac{C_{\up\up\up} + 3C_{\up\up\dn}}{2} 
	+	\frac{C_{\up\up\dn} - C_{\up\up\up}}{4}
		\widetilde{u}_\text{kag} (\beta J_{bb})
					\nonumber\\&{}~~
	+	\left( \frac{U_{\up\up\dn} - U_{\up\up\up}}{4} \right)^2
		\frac{\beta^2}{\beta_\text{kag} ^2}
		\widetilde{c}_\text{kag} (\beta J_{bb})
	,
	\end{align}
where $\widetilde{u}_\text{kag}$ is the energy per unit cell of the kagome lattice Ising ferromagnet (in units of the kagome coupling) and $\widetilde{c}_\text{kag}$ is the heat capacity per unit cell of the kagome lattice.\cite{syozi51,kasteleyn1963,fisher1966,horiguchi1992}
The entropy per unit cell is $s=f + \beta u$.  The ground-state entropy can be calculated by taking the limit $\beta\rightarrow \infty$ and observing that $U_j(\{\sigma_b\})$ and $C_j(\{\sigma_b\})$ are dominated by the lowest energy levels of the hexamer.


\subsection{Zero-temperature phase diagram and ground-state properties of each phase\label{groundstate}}
\begin{figure}[t]
\psfrag{Jaax}{$J_{aa}^X$}
\psfrag{Jaaz}{$J_{aa}^Z$}
\psfrag{Jab=-1}{$$}
\psfrag{I1}{I(3/2)} 
\psfrag{I2}{I(1/2)} 
\includegraphics[width=0.95\columnwidth]{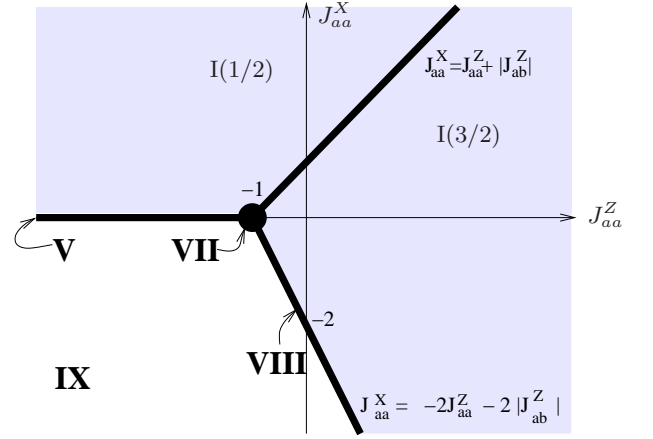}
\caption{(Color online) Exact $T=0$ phase diagram of the TKL
XXZ-Ising model in the $(J_{aa}^Z, J_{aa}^X)$ plane, with $|J_{ab}^Z|=1$ as the unit of energy.
The black lines indicate phase boundaries.  
Phases I(1/2) and I(3/2) are described in the text; they are ferromagnetic if 
$J_{ab}^Z>0$ and ferrimagnetic if $J_{ab}^Z<0$.
}
\label{f:ising-heisenberg-phase-diagram}
\end{figure}

\begin{figure}[t]
\psfrag{Jaax}{$J_{aa}^X$}
\psfrag{Jaaz}{$J_{aa}^Z$}
\psfrag{T}{$T$}
\includegraphics[width=0.95\columnwidth]{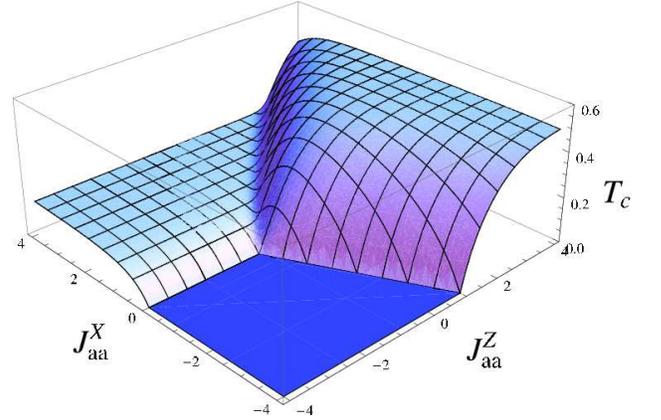}
\caption{(Color online)
Finite-temperature phase diagram of the TKL XXZ-Ising model, with $|J_{ab}^Z|=1$ as the unit of energy.
In the dark blue region of parameter space (phase IX), the system is disordered even at zero temperature.  The straight lines in Fig.~\ref{f:ising-heisenberg-phase-diagram} corresponding to phases V and VIII turn into critical surfaces in Fig.~\ref{f:3dphase}.  However, the boundary line between I(1/2) and I(3/2) in Fig.~\ref{f:ising-heisenberg-phase-diagram} is hidden underneath the critical surface in  Fig.~\ref{f:3dphase}.  
}
\label{f:3dphase}
\end{figure}

In the absence of $h$,  the partition function is
invariant under a change of sign of $J_{ab}^Z$ (because this can be gauged
away by redefining the Ising $b$-spins), so the key physics 
is independent of the sign of $J_{ab}^Z$ . However, because of the frustrated geometry of the trimer, it is {\em not} 
invariant under a change
of the sign of $J_{aa}^X$.

The behavior of the kagome Ising ferromagnet is determined by the dimensionless coupling $K= \beta J_{bb}$. For  $T=0$, we need the value $K_0$ which is the limiting value of $K$ as $\beta$ goes to  infinity. In that limit, each $a$-trimer is restricted to its ground state for a given configuration of the surrounding $b$-spins, and Eq.~\eqref{e:a-and-jbb} becomes
\begin{equation}
        \beta J_{bb} = \frac{1}{4} [\ln \frac{D_1}{D_2} + \beta(E_0(\up\up\dn)-E_0(\up\up\up))]
\end{equation}
where $E_0(\up\up\up)$ and $E_0(\up\up\dn)$ are the $a$-trimer ground state energies for the two relevant configurations of the $b$-spins, and $D_1$ and $D_2$ are their respective degeneracies. Provided $E_0(\up\up\dn)$ is greater than $E_0(\up\up\up$), then clearly $K_0$ goes to infinity, and the $b$-spins will have perfect ferromagnetic order. On the other hand, 
if ${E_0(\up\up\dn)} = {E_0(\up\up\up)}$,
$K_0$ will be finite but can still be $>0$ if $D_1>D_2$. In this case, whether the $b$-spins have LRO depends on whether $K_0$ exceeds the critical coupling of the kagome Ising model. Note that here the coupling between the $b$-spins is caused by maximizing the ground state degeneracy.

By studying the energy levels $E_0(\up\up\dn)$ and $E_0(\up\up\up)$ for different combinations of parameters, we can work out the entire zero-temperature phase
diagram in the $(J_{aa}^Z,J_{aa}^X)$ plane, which is shown in
Fig.~\ref{f:ising-heisenberg-phase-diagram}.  Furthermore, the values of parameters at which the energy levels
$E_j(\up\up\up)$ or $E_j(\dn\up\up)$ undergo level crossings are also of importance, as it changes the value of $D_1$ and/or $D_2$. We label the phases I, V, VII, VIII, IX for consistency with           
our previous work\cite{lohyaocarlson2007}.        
Table~\ref{t:phases} shows the residual entropy of each of these phases, calculated using the approach described in the previous section.

\begin{figure}[t]
\includegraphics[width=0.85\columnwidth]{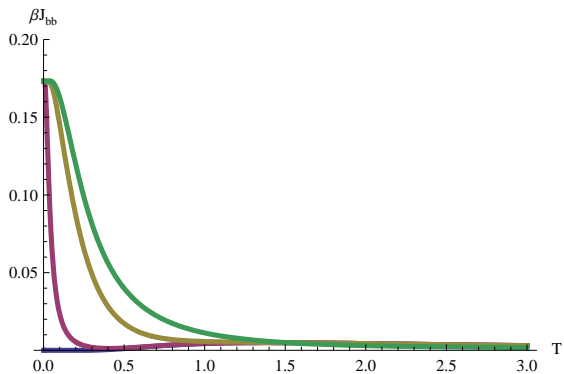}
\caption{(Color online) The dimensionless effective Ising two-spin coupling
  between adjacent $b$-spins without magnetic field, $\beta J_{bb}$, as a function of temperature $T$. There are four values of the transverse coupling $J_{aa}^X=0, -0.1, -1, -10$ (bottom to top).  Other parameters are $J_{aa}^Z=-3$ and $J_{ab}^Z=\pm 1$.
}
\label{f:betaJbb}
\end{figure}

\begin{table}[!h]
\begin{ruledtabular}
\begin{tabular}{cccccc} 
Phase & $\beta J_{bb}$               & $s_0$            & $s_0/9$    & $\xi$      \\ \hline
I    & $\infty$                      &  $0$             &  $0$       & 0 (LRO)        \\
V    & $0$                           &  $\ln 72=4.2767$ &  $0.47519$ & $0$        \\
VII    & $\frac{1}{4}\ln\frac{4}{3}$ & $4.4368$         & $0.49297$  & finite     \\
VIII   & $\frac{1}{4}\ln 3$          & $2.8918$         & $0.32131$  & finite     \\
IX  & $\frac{1}{4}\ln 2$             & $2.5258$         & $0.28064$  & finite     \\
\end{tabular}
\end{ruledtabular}
\caption{Properties of the various $T=0$ phases marked on the phase diagram in Fig.~\ref{f:ising-heisenberg-phase-diagram}.  Here, $\beta J_{bb}$ is the effective Ising coupling between $b$-spins, $s_0$ is the entropy of the entire system per TKL unit cell, $s_0/9$ is the entropy per site, and $\xi$ is the correlation length of the $b$-spins.  Comparing $\beta J_{bb}$ to the $\beta J_c$ of the kagome lattice shows that only phase I has long-range order.
\label{t:phases}}
\end{table}

Note that the entropies satisfy the inequalities $S_{V}, S_{VII},
S_{VIII} > S_{IX},S_{I}$ 
and $S_{VII} > S_{V}, S_{VIII}$.  This agrees with
the intuition that when the system is tuned to 
a phase transition line or point,  the system is able to access states from both adjacent
phases, and therefore the entropy is higher than that of the
surrounding phases.

Fig.~\ref{f:3dphase} shows the entire finite temperature phase diagram 
as a function of $J_{aa}^Z/|J_{ab}^Z|$, $J_{aa}^X/|J_{ab}^Z|$,
and $T/|J_{ab}^Z|$.
Apart from the I(3/2) to I(1/2) ground state transition, which
corresponds only to a change in the local physics, it is clear that all of the phase transitions survive at finite temperature.  
Not surprisingly, as temperature is increased, the
disordered phase becomes a larger part of the phase diagram.

Based on experimental data on 
the TKL magnets
$\mbox{Cu}_{9}\mbox{X}_2(\mbox{cpa})_{6}\cdot x\mbox{H}_2\mbox{O}$,~\cite{gonzalez93,maruti94, mekata98}, 
it is sensible to make the following assumptions:  
$J_{aa}^Z$ and $J_{aa}^X$ are antiferromagnetic ($J_{aa}^Z,J_{aa}^X<0$),
and $|J_{aa}^Z| \gg |J_{ab}^Z|$.
At zero applied field, this would put these materials in the disordered phase.

\subsection{Physical explanation}

We now discuss the 
various phases and the transitions between them. Phase I has ${E_0(\up\up\dn)}>E_0(\up\up\up)$,
as is shown in the right side of Fig.~\ref{f:energy-levels-of-a-trimer},
and therefore $K_0=K(T\rightarrow 0)$ 
is infinite (see Sec.~\ref{groundstate}). 
In this case the $b$-spins are perfectly ferromagnetically ordered at $T=0$. This ferromagnetic phase of the $b$-spins is further subdivided into two $T=0$ phases by the line $J_{aa}^X=J_{aa}^Z+|J_{ab}^Z|$ (see Fig.~\ref{f:ising-heisenberg-phase-diagram}), corresponding to different $a$-trimer configurations. 
In the phase I(3/2), the $a$-trimers are in a unique ground state with total spin value $S=3/2$
and total $z$-component of spin  $S^Z=3/2$.
In the phase I(1/2), the $a$-trimers are in a unique ground state with 
total spin value $S=3/2$ and total $z$-component of spin $S^Z=1/2$.  
On the phase boundary, these two states become degenerate, and each $a$-trimer is two-fold degenerate, 
with an associated $T=0$ entropy of $\ln 2$.    
At finite temperature, the transition from I(3/2) to I(1/2) becomes a 
crossover, since 
the difference between phases 
I(3/2) and I(1/2) is the local configuration of each $a$-trimer.
This crossover, which is evident in the region $J_{aa}^X=J_{aa}^Z+|J_{ab}^Z|$
in Fig.~\ref{f:3dphase}, should be accompanied by a relatively sharp 
peak in the entropy near the $T=0$ phase boundary. From the figure, it can also be seen that the I(3/2) phase is more robust than the I(1/2) phase, in that it has a higher transition temperature. This can be qualitatively understood as follows. In phase I, the effective coupling $J_{bb}$ at $T=0$ is determined by $E_0(\up\up\dn)-E_0(\up\up\up)$. 
Far into phase I (large $|J_{aa}|$), the $a$-trimer ground state is locked in its $J_{ab}^Z=0$ configuration, and $E_0(\up\up\dn)-E_0(\up\up\up)$ can be calculated by perturbation theory to be $|J_{ab}^Z|$ for I(3/2)
and  $\frac{2}{3} |J_{ab}^Z|$ for I(1/2). Thus, I(3/2) has a larger $J_{bb}$ than I(1/2) and therefore a larger transition temperature.
 

Let us consider going from phase I(1/2) to phase IX across the phase boundary V. Figure \ref{f:energy-levels-of-a-trimer} shows the eight energy levels of an $a$-trimer, for the two symmetry-distinct configurations of the $b$-spins ($\up\up\up$ and $\dn\up\up$), with $J_{aa}^Z=-3$ and $J_{ab}^Z=\pm 1$.
The most important feature of the graph is how $E_0(\up\up\dn)$ compares with $E_0(\up\up\up)$, 
along with their respective degeneracies.
Starting from phase I(1/2), as $J_{aa}^X$ is decreased towards $0$, the difference between $E_0(\up\up\dn)$ and $E_0(\up\up\up)$ decreases until they become equal at the full Ising limit (phase V), $J_{aa}^X=0$.   
The ground state of the $a$-trimer is triply
degenerate in this limit, regardless of the configuration of the $b$-spins.  
Because of this, the effective coupling between $b$-spins goes to zero, $K_0=0$. 
This was noted in our previous paper\cite{lohyaocarlson2007}
on the Ising limit of the TKL model, 
along with the corresponding ground state entropy of 
$\ln 72$ 
per TKL unit cell in the Ising limit.

Now turn on an antiferromagnetic transverse coupling
$J_{aa}^X<0$ to go into phase IX. The lowest energy remains independent of the $b$-spin configurations. 
Notice, though, that the lowest eigenvalue of the $(\up\up\up)$ subspace 
is doubly
degenerate, whereas the lowest eigenvalue of the 
$(\dn\up\up)$ subspace is nondegenerate.
In the $(\up\up\up)$ subspace of the $b$ spins, 
the lowest $a$-trimer energy state in this region has $S=1/2$, $S^Z=1/2$, and there are two independent states with these quantum numbers. 
They are degenerate because the surrounding ($\up\up\up$) $b$-spins 
couple to the $a$-spins through $S^Z$ only. 
When the $b$-spins are in the $(\dn\up\up)$ configuration, this degeneracy is broken. This means
that at zero temperature, the $(\up\up\up)$ configuration of the
$b$-spins is twice as likely to occur as the $(\dn\up\up)$ 
configuration. 
The zero temperature effective coupling constant between $b$-spins is therefore 
$K_0= \frac{1}{4}\ln 2$.

To get a fuller picture of what is happening, we plot the
effective coupling $\beta J_{bb}$ as a function of temperature
in Fig.~\ref{f:betaJbb}. 
We see that 
as long as quantum fluctuations are present (i.e. $J_{aa}^X$ is finite),
the value of $\beta J_{bb}$ increases monotonically as $T$ is lowered, 
approaching  a constant
value $\frac{1}{4}\ln 2 = 0.173287...$ as $T \rightarrow 0$.
This value is less than the
critical coupling of the kagome Ising ferromagnet, $\beta J_{bb}^{c} = \frac{1}{4}
\ln (3 + \sqrt{12}) = 0.466566...$.  
Therefore phase IX is disordered at all temperatures.
The entropy in Phase IX is 
 $2.5258...$ per unit cell.
It exists precisely at $h=0$ and $J_{aa}^X = J_{aa}^Z < -1$.  
Note that in deviating from the Ising antiferromagnetic limit
(i.e. Phase V) by adding quantum fluctuations, the
entropy decreases.  

Next we consider crossing the phase boundary from I(3/2) to IX through VIII. Along the line VIII, we again have $E_0(\up\up\dn)=E_0(\up\up\up)$. The difference from IX,  however, is that when the $b$-spins are in the $(\up\up\up)$ configuration, the $S=3/2,~S^z=3/2$ state of the $a$-trimer is also degenerate with the two  $S=1/2,~S^Z=1/2$ states. 
This leads to a larger value of $K_0 = \frac{1}{4}\ln 3$, and hence a longer $T=0$ correlation length, than phase IX. Nevertheless, this value of $K_0$ is still below that of the critical value of the kagome Ising model, and the ground state remains disordered.

If we are in phase IX close to the line phase VIII, the $a$-trimer ground state when the surrounding $b$-spins are all up is doubly degenerate, with a small gap to the $S=3/2$ state. If the temperature is now increased, then this low-lying excited state will have non-zero Boltzmann weight, and hence 
the value of $K$ will first increase 
before decreasing again at higher temperature due to the effects of thermal disordering. 
Thus, the change of $K$ and hence the $b$-spin correlation length with temperature will be non-monotonic, first increasing with temperature before decreasing. This is a manifestation of the order-by-disorder mechanism commonly seen in frustrated systems with residual ground state entropy.


%

\section{Finite Magnetic Field \label{finiteh}}
\subsection{Exact mapping to the kagome Ising model with three-spin interactions \label{sec:kagome-three-spin}}  


We now consider the XXZ-Ising model on the TKL 
in the presence of finite magnetic field.  
This model is defined in Eq.~\eqref{e:tkl-quantum-heisenberg-hamiltonian}.
Note that we consider applied field parallel to the
axis of the Ising spins on the $b$ sublattice,
so that the $b$ spins remain classical in their behavior.
In the presence of finite field $h$, spins on the $a$-sublattice
can still be integrated out, yielding an effective
model in terms of the $b$ spins only, which
reside on a kagome lattice.  
However, since the original model for $h\ne0$ has
explicitly broken time-reversal symmetry, it is
necessary to allow for the possibility of $3$-spin
couplings in the effective model for the $b$ spins. 
The effective Hamiltonian of the $b$ spins $H_\text{eff}(S_b)$ is therefore of the form
 \begin{equation}
        H_\text{eff}(S_b)=- J_{bb}\sum_{\mean{i,j}} \sigma_{bi} \sigma_{bj}
        -J_{bbb}\sum_{i,j,k} \sigma_{bi} \sigma_{bj} \sigma_{bk} -h_\text{eff}\sum_i \sigma_{bi},
\end{equation}
which is an Ising model on the kagome lattice
with three-spin interactions occuring within each $b$-trimer.  (A brief numerical study of a kagome Ising model with three-spin interactions can be found in Ref.~\onlinecite{beirl1993}.)
\begin{figure}[t]
\includegraphics[width=0.85\columnwidth]{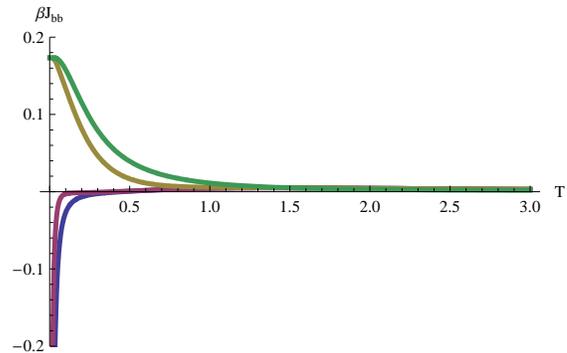}
\caption{(Color online) The dimensionless effective Ising two-spin coupling
  between adjacent $b$-spins with a finite magnetic field $h=0.1$, $\beta J_{bb}$, as a function of temperature $T$,
  for four values of the transverse coupling $J_{aa}^X=0, -0.1, -1, -10$ (bottom to top).  Other parameters are $J_{aa}^Z=-3$ and $J_{ab}^Z=\pm 1$.
}
\label{hbetaJbb}
\end{figure}

In the same manner in which we proceeded in Sec.~\ref{sec:exact}, 
we can integrate out the $a$ spins to form an effective model for the $b$ spins.
       \begin{align}
        Z &=\Tr \exp (-\beta \hat H) \\
        & =\sum_{\{S_b\}} \Tr_{S_{a1},S_{a2},S_{a3}} \exp \left[-\beta \hat H (S_b,S_a,h)\right] \\
        & =\sum_{\{S_b\}} e^{-\beta \hat H_\text{eff}(S_b)}.
        \label{e:partition-function-factorization-finite-field}
        \end{align}
In the presence of an applied field, time-reversal symmetry is explicitly broken,
and so there are four distinct terms in the trace.
$Z({S_{b1},S_{b2},S_{b3}})$ for each $b$-trimer must be matched to the 
new effective model as follows (remembering to divide $h_\text{eff}$ by two because it is shared between two adjacent hexamers): 
  \begin{align}
  &  Z(\up\up\up,h)=Z_a \exp\left[ \beta(3J_{bb}+J_{bbb}+\frac{3h_\text{eff}}{2})\right], \\
  &  Z(\up\up\dn,h)=Z_a \exp \left[\beta(-J_{bb}-J_{bbb}+\frac{h_\text{eff}}{2})\right], \\
  &  Z(\up\dn\dn,h)=Z_a \exp \left[\beta(-J_{bb}+J_{bbb}-\frac{h_\text{eff}}{2})\right], \\
  &  Z(\dn\dn\dn,h)=Z_a \exp \left[\beta(3J_{bb}-J_{bbb}-\frac{3h_\text{eff}}{2})\right]. 
   \label{zh}
  \end{align}
From these equations, we find
   \begin{align}
     &\beta J_{bb} = \frac{1}{8}\ln \frac{Z(\up\up\up,h)Z(\dn\dn\dn,h)}{Z(\up\up\dn,h)Z(\up\dn\dn,h)}, \\
     &\beta J_{bbb}=\frac{1}{8}\ln  \frac{Z(\up\up\up,h)Z(\up\dn\dn,h)^3}{Z(\dn\dn\dn,h)Z(\up\up\dn,h)^3},\\
     &\beta h_\text{eff} =\frac{1}{4} \ln \frac{Z(\up\up\up,h)Z(\up\up\dn,h)}{Z(\dn\dn\dn,h)Z(\up\dn\dn),h},\\
     &Z_{a} =\exp(f_a)= \left[Z(\up\up\up,h)Z(\dn\dn\dn,h)\right]^{1/8}  \\
&\hspace{.5in}      \times \left[Z(\up\up\dn,h)Z(\up\dn\dn,h)\right]^{3/8}.
    \end{align}
The values of $Z({S_{b1},S_{b2},S_{b3}})$ are given by
Eqs.~\eqref{e:zuuu-and-zduu-1}--\eqref{e:zuuu-and-zduu}.  For
$h\rightarrow 0$, we find that $J_{bbb}\rightarrow 0$ and
$h_\text{eff}\rightarrow 0$, and these equations reduce to our
previous results, Eq.\eqref{e:a-and-jbb}).  
In Fig.~\ref{hbetaJbb}, we
show how the external field $h$ changes the effective coupling
$J_{bb}$ as a function of temperature and $J_{aa}^X$. We see that a
small field can change the sign of $J_{bb}$ when $J_{aa}^X$ is weak.

\begin{figure}[t]
{\centering
  \subfigure[~$J_{bb}>0$\label{kagomephase2}]{\resizebox*{0.6\columnwidth}{!}{\includegraphics{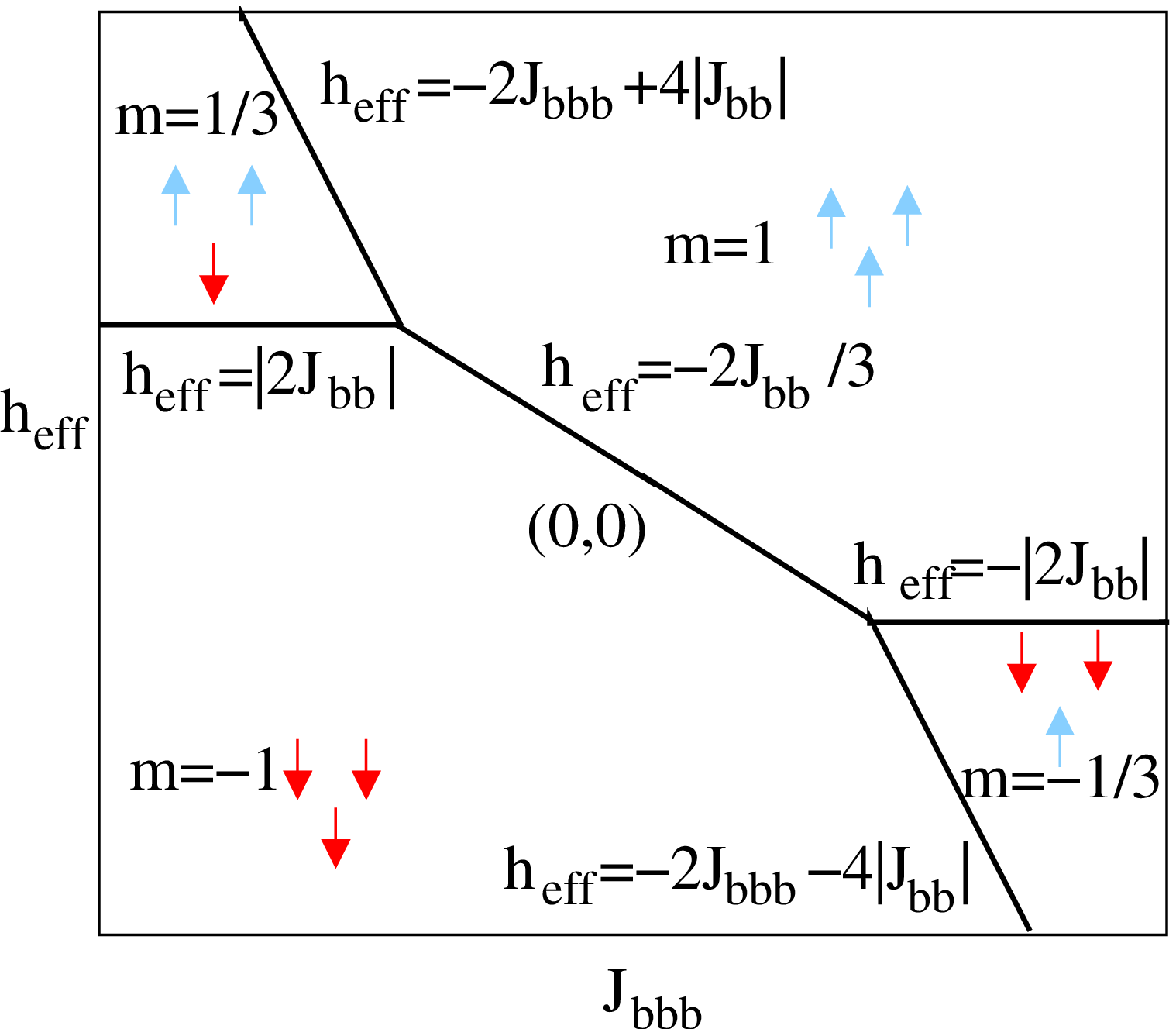}}}
  \subfigure[~$J_{bb}=0$\label{kagomephase0}]{\resizebox*{0.6\columnwidth}{!}{\includegraphics{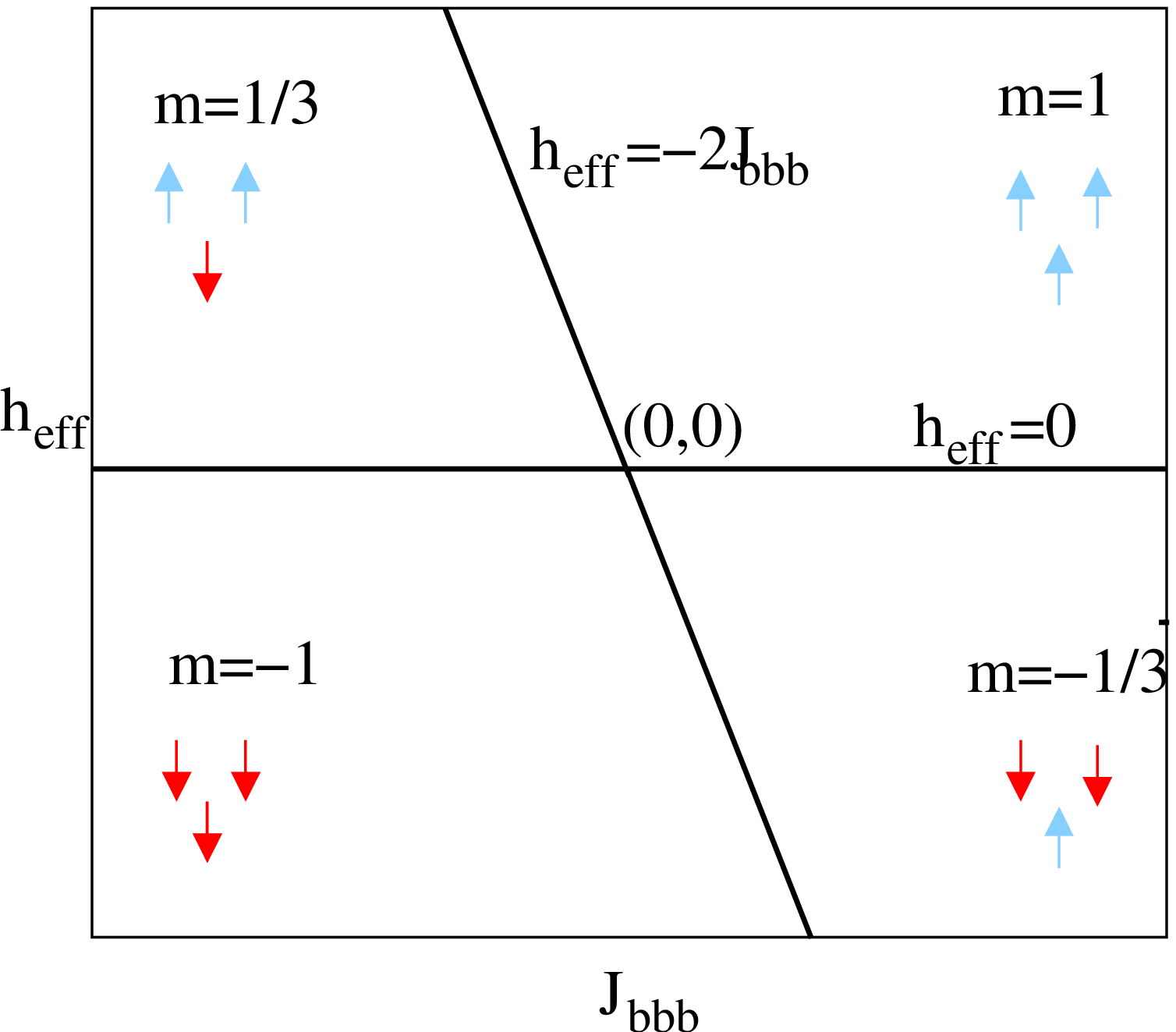}}}
  \subfigure[~$J_{bb}<0$\label{kagomephase1}]{\resizebox*{0.6\columnwidth}{!}{\includegraphics{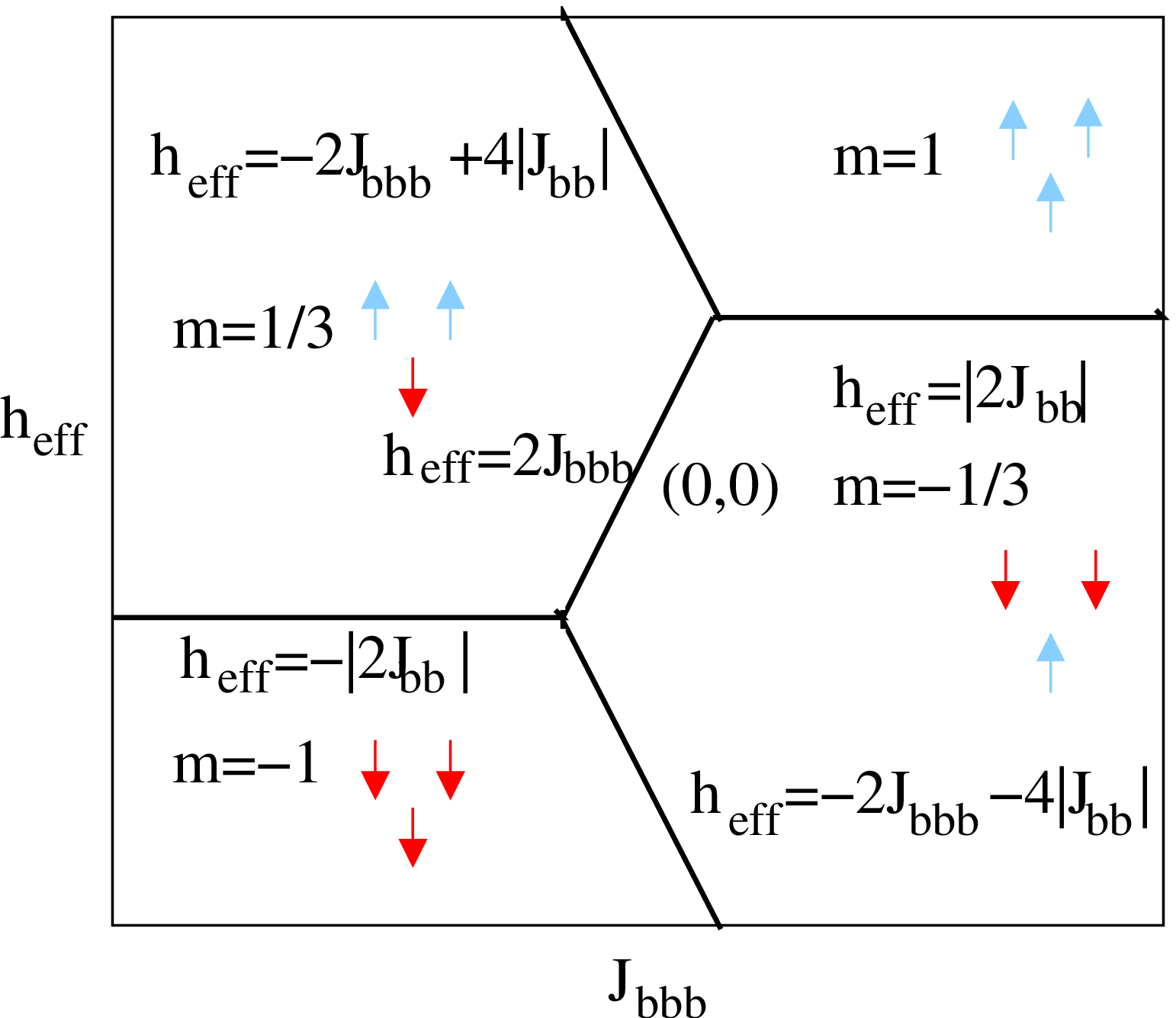}}} \par}
\caption{(Color online) Ground state phase diagram of kagome Ising model 
with three-spin interactions.  (a) Ferromagnetic $J_{bb}>0$. (b) $J_{bb}=0$. (c) Antiferromagnetic $J_{bb}<0$. $m$ is the average magnetization per site.}  
\label{kagomephase}
\end{figure}



\subsection{Ground state phase diagram of the 3-spin kagome Ising model\label{sec:3-spin}}
We now have an exact mapping from the TKL in finite field to the kagome Ising model
with the three-spin interaction term $J_{bbb}$.
As we will see in the next section, the full ground state phase diagram of the 
original model will include phase transitions which only involve local $a$-spin
configurations.  However, we can develop some intuition about the long-range
physics by studying the effective model for the $b$ spins. 
Since this is now a classical model, the zero temperature phase diagram can be found in the usual manner
by calculating the energy levels of each unit cell in order to determine the ground state:
\begin{align}
   & E_k(\up\up\up)=-3J_{bb}-J_{bbb}-\frac{3}{2}h_\text{eff}, \\
   & E_k(\up\up\dn)= J_{bb}+J_{bbb}-\frac{1}{2}h_\text{eff}, \\
   & E_k(\up\dn\dn)= J_{bb}-J_{bbb}+\frac{1}{2}h_\text{eff}, \\
   & E_k(\dn\dn\dn)=-3J_{bb}+J_{bbb}+\frac{3}{2}h_\text{eff}.
\end{align}
Representative ground state phase diagrams for this model are
shown in Fig.~\ref{kagomephase}.
At large enough effective field strength $|h_\text{eff}|$, 
the system goes into a saturated ferromagnetic phase.
However, large magnitude of the three-spin interaction $|J_{bbb}|$ 
can frustrate this effect.  
In large parts of the phase diagram, we find that the ground state
of a $b$-trimer is in the $(\up\dn\dn)$ state (or its time-reversed
counterpart).  
When  $E_k(\up\up\dn)$  ($E_k(\up\dn\dn)$) is favored, then understanding the
character of the macroscopic ground state is reduced to the
problem of enumerating the ways of tiling the kagome plane with
one down (up) spin per kagome triangle.  
As was shown in Ref.~\onlinecite{moessner2000},
this is equivalent to placing dimers on the
bonds of a honeycomb lattice. 
(See Fig.~9 of Ref.~\onlinecite{lohyaocarlson2007}.)
In this phase, the correlation function
$\left< \sigma_b(\mathbf{0})\sigma_b(\mathbf{r}) \right>$ is equivalent to the
dimer-dimer correlation function, which has been shown\cite{moessner2000}  to be a power
law, $1/r^2$.

In the presence of the three-spin coupling, the topology of the phase
diagram is dependent upon the sign of $J_{bb}$.  
When $J_{bb}>0$, phase transitions between the 
two oppositely polarized dimer phases ({\em i.e.} from 
$(\up\up\dn)$ to $(\up\dn\dn)$)
are forbidden, and the dimer phases are separated by
the saturated ferromagnetic phases.
However, it is possible to go directly from one saturated
ferromagnetic phase to its time-reversed counterpart.
When  $J_{bb}<0$, then phase transitions
directly from one dimer phase into its
time-reversed counterpart are allowed, whereas 
phase transitions between the two saturated phases are not allowed.

\subsection{Full $T=0$ phase diagram of the XXZ-Ising model in finite magnetic field \label{sec:xxz-ising-finite-field}}


The full $T=0$ phase diagram can be obtained by adapting the method described in Sec.~IV-A of Ref.~\onlinecite{lohyaocarlson2007}.  For every configuration of the three $b$-spins in a hexamer, we have tabulated the energies of the eight `internal' states of the hexamer in Table~\ref{t:hexamer-energy-levels}.  That is, we know the energies of the 64 states of the hexamer (including $a$- and $b$-spins).  The pattern of ground states within this 64-element matrix (for a given parameter set) allows us to determine the global ground state of the 
system.\footnote{In general, knowledge of the \emph{finite-temperature thermodynamics} of a \emph{subsystem} does not necessarily allow one to deduce the thermodynamic behavior of the entire system.  It is fortunate that the TKL is simple enough such that knowledge of the \emph{ground state} of a single hexamer allows one to infer the ground state of the entire system.}
Table \ref{t:numerical-example} illustrates this procedure for a particular choice of parameters.

\begin{table}[htbp]
\begin{align*} 
%
%
%
%
\begin{array}{l|llllllll}
  & \text{E1} & \text{E2} & \text{E3} & \text{E4} & \text{E5} & \text{E6} & \text{E7} & \text{E8} \\
  \hline
 \uparrow \uparrow \uparrow  & 0.75 & -2.3 & -2.3 & 0.75 & 0.75 & 0.75 & -2.3 & -2.3 \\
 \uparrow \uparrow \downarrow  & -1.8 & 0.25 & 2.3 & -1.8 & -1.2 & 1.7 & \fbox{-2.5} & 0.98 \\
 \uparrow \downarrow \downarrow  & -0.25 & 3.8 & -0.25 & -2.3 & -1.7 & 1.2 & -0.98 & 2.5 \\
 \downarrow \downarrow \downarrow  & 3.2 & 0.25 & 0.25 & 5.2 & -0.75 & 1.2 & -1.8 & -1.8
\end{array}
%
%
%
%
\end{align*}
\caption{Energies of hexamer states for the parameters $J_{aa}^Z=J_{aa}^X=-2$, $J_{ab}^Z=-1$, $h=2$ to two significant figures.  The rows correspond to different (classical) configurations of the $b$-spins ($\up\up\dn$ represents the three configurations $\dn\up\up$, $\up\dn\up$, and $\up\up\dn$), while the columns correspond to different (quantum) states of the $a$-spins.  The ground state of the hexamer has energy $-2.5$ (indicated by the box), and it has the $b$-spins in one of the three configurations  $\dn\up\up$, $\up\dn\up$, or $\up\up\dn$.  This implies that the ground state of the system is the honeycomb dimer phase (phase IV).
\label{t:numerical-example}
}
\end{table}

In principle, we could, by comparing the analytic expressions for the energy levels (Table~\ref{t:hexamer-energy-levels}), derive analytic expressions for the phase boundaries.  In this paper we will just present a phase diagram obtained by taking a grid of closely-spaced parameter points and finding the hexamer ground state(s) numerically for each parameter point.

\begin{figure}[htb]
{\centering
  \subfigure[ XXX-Ising case\label{tklheiphase}]{\includegraphics[width=0.95\columnwidth]{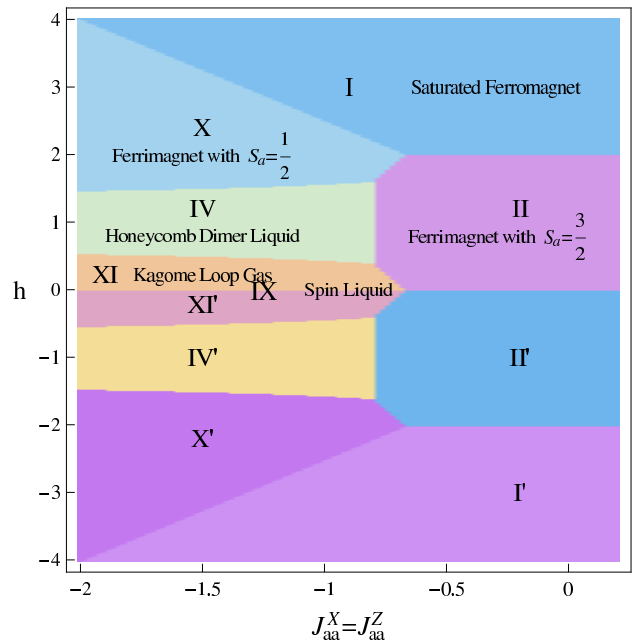}}
  \subfigure[ Ising case\label{tklisingphase}]{\includegraphics[width=0.95\columnwidth]{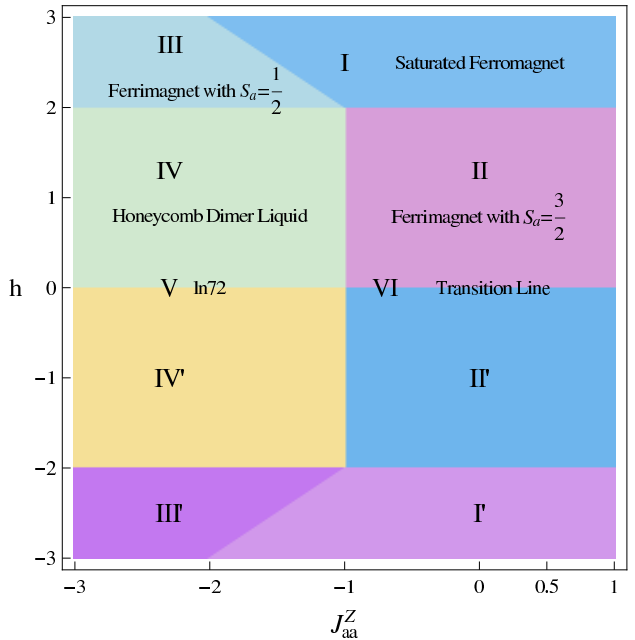}}
}
\caption{(Color online) 
Phase diagrams of the TKL Heisenberg-Ising model for (a) the $a$-spins in the Heisenberg limit, $J_{aa}^X=J_{aa}^Z$,
and (b) the $a$-spins in the Ising limit, $J_{aa}^X=0$.
In both panels, we use $J_{ab}^Z=-1$ and $T=0$.   
}
\label{f:fulltklphase}
\end{figure}

Fig.~\ref{f:fulltklphase} shows the results of this procedure in the limit where the $a$-spins have isotropic
Heisenberg couplings ($J_{aa}^X = J_{aa}^Z$), as well as in the previously studied case of Ising $a$-spins
($J_{aa}^X=0$)  for comparison.
Most of the phase boundaries in Fig.~\ref{tklheiphase}
appear to be straight lines, but four of them are slightly curved.  
The topology of Fig.~\ref{tklheiphase} is more complicated than that of Fig.~\ref{kagomephase},
because there may be distinct $a$-trimer
configurations underlying the same $b$-spin phase and because phase
boundaries in Fig.~\ref{kagomephase} may appear as extended regions of phases here.
The ``primed'' phases are  time-reversed versions of the ``unprimed'' phases.
By examining the energy levels in detail (Table \ref{t:numerical-example}), we have identified the nature of each phase.  

Starting from the bottom of phase II and going through the phase
diagram counterclockwise, the various phases are given as follows.
Phase II is  the I(3/2) phase in a ferrimagnetic arrangement, so
that the $a$-spins and $b$-spins are perfectly aligned
antiferromagnetically due to the coupling $J_{ab}^z$, with the
$a$-spins along the direction of the field $h$. With large enough $h$,
the $b$-spins become all up also, and there is a transition from phase
II into phase I, which is a saturated ferromagnet.  Increasing the
strength of the $a$-trimer Heisenberg coupling sufficiently one reaches
phase X. In this phase, the $b$-spins remain ferromagnetically aligned,
but each $a$-trimer is now in a $S=1/2$, $S^Z=1/2$ state, which is doubly
degenerate.  The entropy per unit cell is therefore $2 \ln 2 = 1.3863...$.  
Lowering the field takes us then into phase IV, which is a critical phase with power-law correlations that is equivalent to close-packed dimers on the honeycomb lattice (see Sec.~\ref{sec:3-spin}).
It is instructive to compare Fig.~\ref{tklisingphase} to Fig.~\ref{tklheiphase} to see how the phase diagram evolves when the transverse coupling $J_{aa}^X$ is introduced: 
phases I, II, and IV
survive quantum fluctuations of the $a$-spins introduced by 
 $J_{aa}^X \ne 0$, although the states of the $a$-trimer are somewhat modified.  


In the next section we give a detailed analysis of phase XI.

\subsection{Phase XI: a ``kagome loop gas''}

\begin{figure}[t]
\includegraphics[width=0.8\columnwidth]{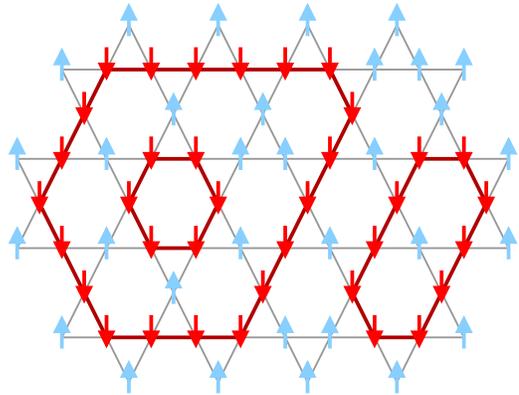}
\caption{(Color online) 
An example of a configuration of $b$-spins in phase XI.  Such configurations can be visualized as loops on the kagome lattice (see text).  The statistical weight of the configuration is proportional to $2^{-l}$, where $l$ is the number of links in the loop.  
}
\label{f:kagomeloops}
\end{figure}

Phase XI is a new and unusual phase.  A typical pattern of hexamer energies for phase XI is shown in Table~\ref{t:weird-energies}.  This means that in phase XI, each $b$-trimer is allowed to contain either 0 down spins or 2 down spins, 
but the latter case 
is less likely
by a factor of 2.  The constraint that each $b$-trimer must contain an even number of down-spins means that the down-spins in the lattice must form 
\emph{nonintersecting closed loops} -- i.e., allowed configurations correspond to \emph{Eulerian subgraphs} of the kagome lattice.  (An Eulerian subgraph is one in which the degree of every vertex is even.)  However, not every Eulerian subgraph corresponds to an allowed configuration of phase XI, because we are not allowed to put three down-spins on the same $b$-trimer.  Figure~\ref{f:kagomeloops} shows an example of an allowed configuration.
In moving through the ``phase space'' of the ground state manifold, 
loop number is not conserved, loops are non-intersecting ({\em i.e.}
they are repulsive with a hard-core interaction), and loop size is not
conserved ({\em i.e.} the loop tension is zero).  This corresponds to a 
non-intersecting loop gas on the kagome lattice.

\begin{table}[htbp]
\begin{align*} 
\begin{array}{l|llllllll}
  & \text{E1} & \text{E2} & \text{E3} & \text{E4} & \text{E5} & \text{E6} & \text{E7} & \text{E8} \\
  \hline
 \uparrow \uparrow \uparrow  & 1.7 & -1.3 & -1.3 & 2.4 & 0.19 & 0.94 & \fbox{-2.1} & \fbox{-2.1}\\
 \uparrow \uparrow \downarrow  & -1.2 & 1.6 & 1.3 & -1.9 & -1.4 & 1.5 & -1.9 & 1.5 \\
 \uparrow \downarrow \downarrow  & -0.81 & 2.4 & 0.69 & \fbox{-2.1} & -1.5 & 1.4 & -1.5 & 1.9 \\
 \downarrow \downarrow \downarrow  & 2.3 & -0.69 & -0.69 & 3.6 & -0.19 & 1.1 & -1.9 & -1.9
\end{array}%
\end{align*}
\caption{Energies of hexamer states, to two significant figures, for the parameters $J_{aa}^Z=J_{aa}^X=-2$, $J_{ab}^Z=-0.25$, $h=2$ (corresponding to phase XI, the ``kagome-loop'' phase).  Boxes indicate ground states.
\label{t:weird-energies}
}
\end{table}

\begin{figure}[t]
{\centering
  \subfigure[$x^0$\label{x00}]{\includegraphics[width=0.31\columnwidth]{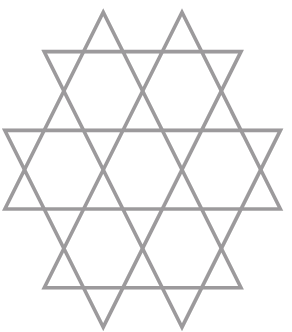}}
  \subfigure[$x^6$\label{x06}]{\includegraphics[width=0.31\columnwidth]{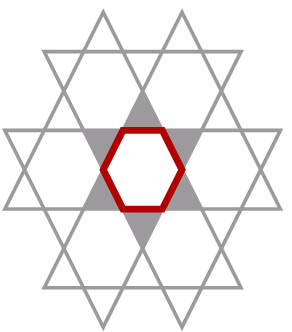}}
  \subfigure[$x^{10}$\label{x10}]{\includegraphics[width=0.31\columnwidth]{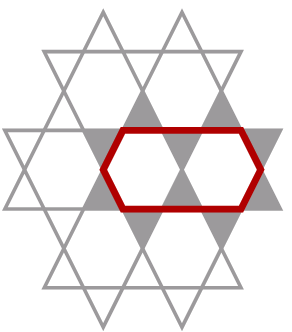}}
  \subfigure[$x^{12}$\label{x12}]{\includegraphics{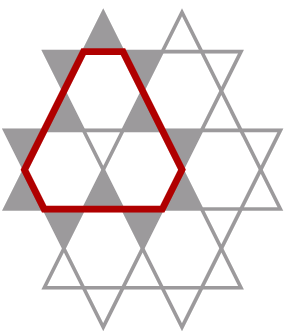}}
  ~~~~
  \subfigure[$x^{12}$\label{x12b}]{\includegraphics{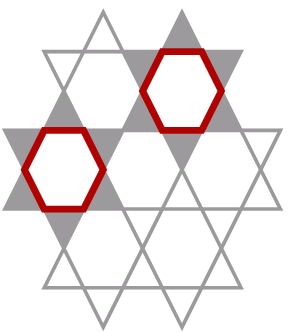}}
}
\caption{(Color online) Diagrams representing terms of up to order $x^{12}$ in the series for the entropy of phase XI (when $h$ is small and positive).  Gray lines represent a background of up-spins on the kagome lattice.  Bold (red) lines represent loops of down-spins.  Shaded triangles represent triangles of $b$-spins in an $\up\dn\dn$ configuration; each shaded triangle corresponds to a factor of $x=\frac{1}{2}$ in the statistical weight.
\label{f:kagome-loop-series}}
\end{figure}

The entropy in phase XI is non-trivial.  First, we present an estimate based on counting loops on the kagome lattice.  This is a series expansion using $x = \frac{  Z(\up\dn\dn)  }{  Z(\up\up\up)  } = \frac{1}{2}$ as a small parameter.
Consider a system with $V$ unit cells.
The most probable configuration of the $b$-spins is the one with
all spins up (Fig.~\ref{x00}).  The next most probable state has a hexagon of
down-spins (Fig.~\ref{x06}).  This state has the same energy (the ground state energy), and comes with a factor of $V$ because the hexagon can be placed anywhere in the lattice, but its statistical weight is smaller by a factor of $x^6$, because there are 6 triangles that are $(\up\dn\dn)$ instead of $(\up\up\up)$.  Continuing in this fashion (Fig.~\ref{x10},\ref{x12},\ref{x12b}), we obtain a series for 
the total number of states:
	\begin{align}
	W &=  2^{2V} \bigg[  1 + V x^{6} + 3V x^{10} 
	\nonumber\\&{}
	+ \left( 2V + \frac{V(V-7)}{2} \right) x^{12}  
	+ \dotso   \bigg],
	\label{e:kagome-loop-degeneracy}
	\end{align}
so the entropy per unit cell is
	\begin{align}
	s &=\lim_{V\rightarrow\infty} \frac{\ln W}{V}
	= 2 \ln 2 + x^6 + 3x^{10} - \frac{3x^{12}}{2} + \dotso .
	\end{align}
When $x=\frac{1}{2}$, including progressively more terms in the series gives the approximations $1.38629$; $1.40192$; $1.40485$; and $1.40448$, suggesting that the entropy per unit cell is $s=1.405(1)$.

The mean magnetization of the $b$-spins can be calculated by a similar series expansion:
	\begin{align}
	m_b &= 3 - 2 \left(
		6 x^6 + 30 x^{10} - 18 x^{12} + \dotso
	\right)
	\end{align}
When $x=\frac{1}{2}$ this series gives $m_b \approx 0.12(1)$.
It seems likely that the true result is an irrational number.  This is in contrast with the honeycomb dimer phase (phase IV), which has rational magnetizations on both sublattices, $m_a=m_b=1/3$.

\begin{figure}[t]
\includegraphics[width=0.5\columnwidth]{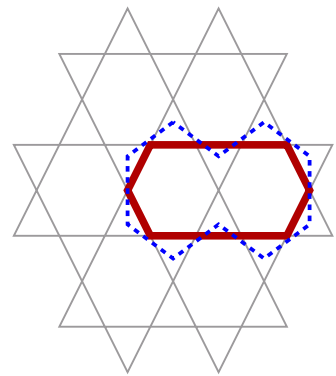}
\caption{(Color online) 
Illustration of the mapping from a configuration of loops on the kagome lattice (bold red lines) to loops on a honeycomb lattice (dashed blue lines).
}
\label{f:kagome-honey-mapping}
\end{figure}

We now present an exact calculation of the entropy of phase XI.
First, note that the kagome lattice is related to a honeycomb lattice in the following way.
Every triangle of the kagome lattice maps to a site of the honeycomb lattice,
and every site (spin) of the kagome lattice maps to a bond on the honeycomb lattice.
Recall that in phase XI, every triangle on the kagome lattice contains either 0 or 2 down spins (call these ``active sites'').
Therefore, every allowed spin configuration on the kagome lattice corresponds to an ``active-bond'' configuration on the honeycomb lattice, in which every site of the corresponding honeycomb lattice must have either 0 or 2 active bonds adjacent to it.  (See Fig.~\ref{f:kagome-honey-mapping}.) 

The higher-order terms in the series for the \emph{degeneracy} of phase XI (Eq.~\eqref{e:kagome-loop-degeneracy}) 
can be seen to be identical to the high-temperature series for the \emph{partition function} of the Ising model on the honeycomb lattice:
	\begin{align}
	Z^\text{honey}_\text{Ising} 
	&= (\cosh \beta J^\text{honey}_\text{Ising}) ^ {3V}    2^V  
		\bigg[
				1 + V t^6 + 3V t^{10} + \dotso
	 \bigg]
	\end{align}
when $t=\tanh \beta J = 1/2$ and $V$ is the number of honeycomb unit cells.  The \emph{entropy} of the kagome loop phase is similarly related to the \emph{free energy} of the honeycomb Ising model.  Using the well-known exact integral expression for the latter\cite{kasteleyn1963,fisher1966}, we find that the entropy of phase XI (per unit cell) is
	\begin{align}
	s^\text{TKL}_\text{phase XI} 
	&= f^\text{honey}_\text{Ising} 
	- 3 \ln \cosh  \beta J^\text{honey}_\text{Ising}
	\nonumber\\{}
	&= 1.4053\dots ,
	\end{align}
which agrees with the series result presented earlier.

It is interesting that the kagome loop gas at $T=0$ and $h\neq 0$ can be exactly mapped to a honeycomb Ising model at $T>0$ and $h=0$.


\section{Conclusions \label{conclusions}}
In conclusion, we have studied an XXZ-Ising model on the triangular
kagome lattice, in which spins residing on small triangles 
($a$-trimers) are fully quantum mechanical, and
spins residing on large triangles ($b$-trimers) are
classical Ising spins.  For all fields and temperatures,
we have shown that there is an exact mapping to the kagome
Ising model.   

In the absence of applied field, 
the full Ising limit ({\em i.e.} $J_{aa}^X = 0$)
of the model has a phase transition from an ordered phase (Phase I) for $J_{aa}^Z$
above a critical value $J_{aa}^{Z,\rm{crit}} = -|J_{ab}^Z|$,
to a disordered phase with 
for $J_{aa}^Z<-|J_{ab}^Z|$ which has residual entropy 
per unit cell $\rm{ln}72 = 4.2767...$
(Phase V).  
The introduction of antiferromagnetic quantum fluctuations of the $a$-spins
({\em i.e.} $J_{aa}^X < 0$)
has little effect on the ordered phase, but it
has dramatic effect on the disordered phase,
partially lifting the degeneracy of the ground state
and reducing the entropy by almost a factor of
two to $2.5258...$ per unit cell.
For strong ferromagnetic quantum fluctuations of the $a$-spins
({\em i.e.} $J_{aa}^X > 0$),
a new phase (phase $I(1/2)$)
appears in which the $a$-trimers
have $S=3/2;~S^Z=1/2$.    There is a zero-temperature phase boundary on the XY side of the Heisenberg
line, for $J_{aa}^X = J_{aa}^Z + |J_{ab}^Z|$,
separating this phase from the one with fully polarized
$a$-trimers, phase $I(3/2)$, with $S=3/2;~S^Z=1/2$
within each $a$-trimer.

All of the ground state phase transitions which occur 
in zero field survive at finite temperature,
except for the ground state transition from 
an ordered phase with $S=3/2; S^Z=3/2$ ({\em i.e.} 
Phase $I(3/2)$)
to an ordered phase with
$S=3/2; S^Z=1/2$ ({\em i.e.} 
Phase $I(1/2)$).  Since this transition is entirely about
local physics, it becomes a crossover at finite temperature.
However, its effects are quite evident at finite temperature,
as the $I(3/2)$ phase is more robust, and survives to much higher temperature
than does the $I(1/2)$ phase.

In the presence of applied field, we find that 
the XXZ-Ising model on the TKL
maps to the kagome Ising model in field
with $3$-spin interactions.  
We find exact solutions for the ground state
of this model.  
In the absence of quantum fluctuations, 
but for strongly frustrated interactions,
we have previously shown that the 
Ising TKL has a critical spin liquid ground state
(with power-law correlations) for any weak applied field
(Phase IV), which can be seen through a mapping
to hard core dimer coverings of the honeycomb lattice.
We find that this phase survives the addition
of quantum fluctuations on the $a$-trimers; however, it is then necessary to apply a finite magnetic field $h$ to access this phase (rather than an infinitesimal field, as in the case of the TKL Ising AF).

For antiferromagnetic interactions, we have shown that small (and even infinitesimal) applied fields lead to a ``kagome loop gas'' ground state (phase XI).  We have mapped this to the ferromagnetic honeycomb Ising model at zero field and finite temperature, and hence we have shown that phase XI has an entropy of $1.4053...$ per unit cell.  
It is likely that the $b$-sublattice magnetization is irrational, in contrast to all the other phases studied in the paper.


\acknowledgments
We are grateful to J.~Stre\v{c}ka for pointing out an error in the entropy of Phase IX in 
an earlier version of our manuscript.
This work was supported by Purdue University (D.X.Y.)
and Research Corporation (Y.L.L.). E.W.C. is a Cottrell Scholar of Research Corporation. 

{\em Note added:} As we were preparing this manuscript for submission, we became aware of similar work 
on the zero-field case done by Stre\v{c}ka \emph{et al.}\cite{strecka2008}  

\end{document}